\input harvmac.tex
\input epsf.tex
\parindent=0pt
\parskip=5pt

\hyphenation{satisfying}

\def\IR{{\hbox{{\rm I}\kern-.2em\hbox{\rm R}}}}
\def\IB{{\hbox{{\rm I}\kern-.2em\hbox{\rm B}}}}
\def\IN{{\hbox{{\rm I}\kern-.2em\hbox{\rm N}}}}
\def\IC{\,\,{\hbox{{\rm I}\kern-.59em\hbox{\bf C}}}}
\def\IZ{{\hbox{{\rm Z}\kern-.4em\hbox{\rm Z}}}}
\def\IP{{\hbox{{\rm I}\kern-.2em\hbox{\rm P}}}}
\def\IH{{\hbox{{\rm I}\kern-.4em\hbox{\rm H}}}}
\def\ID{{\hbox{{\rm I}\kern-.2em\hbox{\rm D}}}}
\def\II{{\hbox{\rm I}\kern-.2em\hbox{\rm I}}}

\noblackbox

\leftline{\epsfxsize1.0in\epsfbox{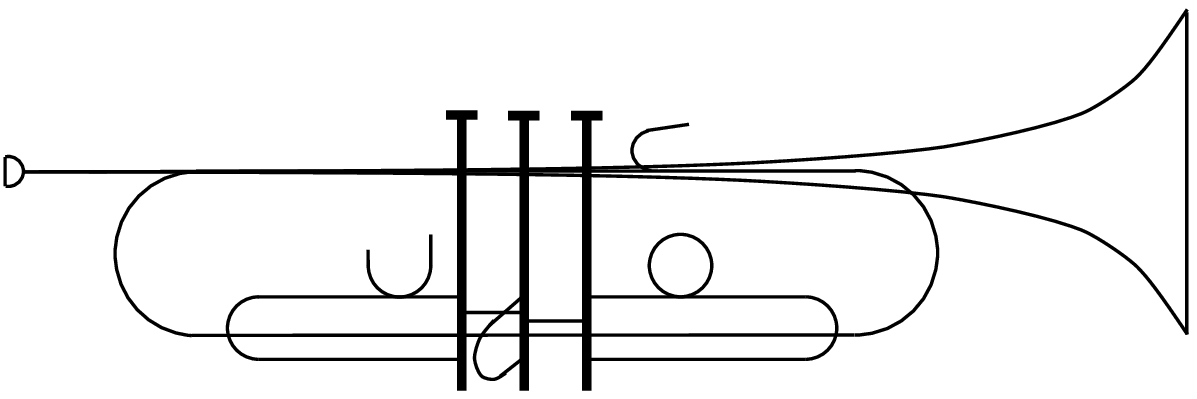}}
\vskip-1.0cm
\Title{\hbox{hep-th/9804200}}
{Superstrings from Supergravity}

\centerline{\bf Clifford V. Johnson$^\flat$}

\bigskip
\bigskip

\vbox{\baselineskip12pt\centerline{\hbox{\it 
Department of Physics}}
\centerline{\hbox{\it University of California}}
\centerline{\hbox{\it Santa Barbara CA 93106, U.S.A.}}}

\footnote{}{\sl email: $^\flat${\tt cvj@itp.ucsb.edu}$\,\,$
On leave from the Dept. of
Physics and Astronomy, University of Kentucky, Lexington KY~40502,
U.S.A.}

\vskip1.0cm
\centerline{\bf Abstract}
\vskip0.7cm
\vbox{\narrower\baselineskip=12pt\noindent
We study the origins of the five ten--dimensional ``matrix
superstring'' theories, supplementing old results with new ones, and
find that they all fit into a unified framework.  In all cases the
matrix definition of the string in the limit of vanishingly small
coupling is a trivial 1+1 dimensional infra--red fixed point (an
orbifold conformal field theory) characterized uniquely by matrix
versions of the appropriate Green--Schwarz action. The Fock space of
the matrix string is built out of winding T--dual strings. There is an
associated dual supergravity description in terms of the near horizon
geometry of the fundamental string solution of those T--dual
strings. The singularity at their core is related to the orbifold
target space in the matrix theory. At intermediate coupling, for the
IIB and $SO(32)$ systems, the matrix string description is in terms of
non--trivial 2+1 dimensional fixed points. Their supergravity duals
involve Anti de--Sitter space (or an orbifold thereof) and are
well--defined everywhere, providing a complete description of the
fixed point theory.  In the case of the type~IIB system, the two extra
organizational dimensions normally found in F--theory appear here as
well. The fact that they are non--dynamical has a natural
interpretation in terms of holography.}

\Date{29th April 1998}
\baselineskip13pt
\lref\dbranes{J.~Dai, R.~G.~Leigh and J.~Polchinski, 
{\sl `New Connections Between String Theories'}, Mod.~Phys.~Lett.
{\bf A4} (1989) 2073\semi P.~Ho\u{r}ava, {\sl `Background Duality of
Open String Models'}, Phys. Lett. {\bf B231} (1989) 251\semi
R.~G.~Leigh, {\sl `Dirac--Born--Infeld Action from Dirichlet Sigma
Model'}, Mod.~Phys.~Lett. {\bf A4} (1989) 2767\semi J.~Polchinski,
{\sl `Combinatorics Of Boundaries in String Theory'}, Phys.~Rev.~D50
(1994) 6041, hep-th/9407031.}

\lref\orientifolds{A. Sagnotti, in {\sl `Non--Perturbative Quantum
 Field Theory'}, Eds. G. Mack {\it et. al.} (Pergammon Press, 1988),
p521\semi V. Periwal, unpublished\semi J. Govaerts, Phys. Lett. {\bf
B220} (1989) 77\semi P. Hor\u{a}va, {\sl `Strings on World Sheet
Orbifolds'}, Nucl. Phys. {\bf B327} (1989) 461.}
\lref\nsfivebrane{A. Strominger,  {\sl `Heterotic Solitons'} Nucl. Phys. 
{\bf B343}, (1990) 167; 
{\it Erratum: ibid.}, {\bf 353} (1991) 565.}
\lref\robin{R. W. Allen, I. Jack and D. R. T. Jones, 
{\sl `Chiral Sigma Models and the Dilaton Beta Function'}, 
Z. Phys. C {\bf 41} 
(1988) 323.}

\lref\banksreview{For a review, with references,
see {\sl `Matrix Theory'}, T. Banks, hep-th/9710231.}
\lref\rutgers{D-E. Diaconescu, J. Gomis, 
{\sl `Matrix Description of Heterotic Theory on K3'}, hep-th/9711105.}
\lref\sen{A. Sen,  {\sl `D0--Branes on $T^n$ and Matrix Theory'}, 
hep-th/9709220.}
\lref\seibergtwo{N. Seiberg, 
{\sl `Why is the Matrix Model Correct?'}, Phys. Rev. Lett. {\bf 79}
 (1997) 3577, hep-th/9710009.}
\lref\seibergthree{N. Seiberg, {\sl `Notes on Theories with 16 
Supercharges'},  hep-th/9705117.}
\lref\paul{ P. S. Aspinwall, {\sl `Some Relationships Between Dualities in 
String Theory'}, Nucl. Phys. Proc. Suppl. {\bf 46} (1996) 30,
hep-th/9508154.}
\lref\john{J. H. Schwarz, 
{\sl `The Power of M Theory'}, Phys. Lett. {\bf B367} (1996) 97,
 hep-th/9510086.}
\lref\juan{J. Maldacena, {\sl `The Large $N$ Limit of Superconformal Field
 Theories and Supergravity'}, hep-th/9711200.}
\lref\juantwo{N. Itzhaki, J. M. Maldacena, J. Sonnenschein 
and S. Yankielowicz, {\sl `Supergravity and The Large N Limit of Theories
With Sixteen Supercharges'}, hep-th/9802042.}
\lref\correspond{S. S. Gubser, I. R. Klebanov, A. M. Polyakov,
 {\sl `Gauge Theory Correlators from Non-Critical String Theory'},
 hep-th/9802109\semi E. Witten, {\sl `Anti De Sitter Space And
 Holography'}, hep-th/9802150.}
\lref\ofer{ O. Aharony, Y. Oz and  Z. Yin,
 {\sl `M--Theory on $AdS_p{\times}S^{11-p}$ and Superconformal Field
 Theories'}, hep-th/9803051\semi E. Halyo, {\sl `Supergravity on
 $AdS_{4/7} \times S^{7/4}$ and M Branes'}, hep-th/9803077.}
\lref\vafa{ C. Vafa, {\sl `Evidence  for F--Theory'}, Nucl. Phys.
 {\bf B469} 403 (1996),  hep-th/9602022.}
\lref\amied{A. Hanany and E. Witten, {\sl `Type IIB Superstrings, 
BPS Monopoles, And Three-Dimensional Gauge Dynamics'}, Nucl. Phys. B492
(1997) 152-190, hep-th/9611230.}
\lref\ofereva{ O. Aharony, M. Berkooz, S. Kachru, N. Seiberg 
and E. Silverstein, {\sl `Matrix Description of Interacting Theories
in Six Dimensions'}, Adv. Theor. Math. Phys. {\bf 1} (1998) 148-157,
hep-th/9707079.}
\lref\edhiggs{E. Witten, {\sl `On The Conformal Field Theory of
 The Higgs Branch'}, JHEP 07 (1997) 003,  hep-th/9707093.}
\lref\stromingervafa{A. Strominger and C. Vafa, {\sl `Microscopic 
Origin of the Bekenstein--Hawking Entropy'}, Phys. Lett. {\bf B379}
(1996) 99, hep-th/9601029.}
\lref\douglasetal{M. R. Douglas, J. Polchinski and A Strominger, 
{\sl `Probing Five-Dimensional Black Holes with D-Branes'}, JHEP
12 (1997) 003, hep-th/9703031.}
\lref\gregsamson{A. Losev, G. Moore and  S. L. Shatashvili,
 {\sl `M\&m's'}, hep-th/9707250.}

\lref\sethisuss{S. Sethi and L. Susskind, {\sl `Rotational 
Invariance in the M(atrix) Formulation of Type IIB Theory'},
 Phys. Lett. {\bf B400} (1997) 265, hep-th/9702101.}

\lref\heteroticcosets{C. V. Johnson, {\sl `Heterotic Coset Models'},
 Mod. Phys. Lett. {\bf A10} (1995) 549, hep-th/9409062\semi {\it
ibid.,} {\sl `Exact Models of Extremal Dyonic 4D Black Hole Solutions
of Heterotic String Theory'}, Phys. Rev.  {\bf D50} (1994) 4032,
hep-th/9403192.}

\lref\diacon{D.--E. Diaconescu and N. Seiberg, 
{\sl `The Coulomb Branch of $(4,4)$ Supersymmetric Field Theories in 
Two Dimensions'}, JHEP, {\bf 07}
 (1997) 001, hep-th/9707158.}
\lref\mikejoeandy{M. R. Douglas, J. Polchinski, and A. Strominger, 
{\sl`Probing
 Five Dimensional Black Holes with D--branes'}, hep-th/9703031.}
\lref\callan{C. G. Callan, J.A. Harvey and A. Strominger, 
{\sl `Supersymmetric String Solitons'}, in
Trieste 1991, proceedings, ``String Theory and Quantum Gravity'',
hep-th/9112030.}
\lref\sjrey{S--J. Rey, in
 {\sl `Superstrings and Particle Theory: Proceedings'}, edited by
   L. Clavelli and B. Harms, (World Scientific, 1990) 291\semi
   S--J. Rey, {\sl `The Confining Phase of Superstrings and Axion Strings'},
 Phys. Rev. {\bf D43} (1991) 526\semi I. Antoniades,
   C. Bachas, J. Ellis and D. Nanopoulos, {\sl `Cosmological String
   Theories and Discrete Inflation'}, Phys. Lett.  {\bf B211} (1988)
   393\semi {\it ibid.,} {\sl `An Expanding Universe in String
   Theory'}, Nucl. Phys. {\bf 328} (1989) 117.}
\lref\robdilaton{R.~C.~Myers, {\sl `New Dimensions for Old Strings'}, 
Phys. Lett. {\bf B199} (1987) 371.}
\lref\ff{B. L. Feigin and D. B. Fuchs, Funct. Anal. Appl. {\bf 16} (1982)
 114, {\it ibid.}, {\bf 17} (1983) 241.}
\lref\gepner{D. Gepner, {\sl `Spacetime Supersymmetry in Compactified 
String Theory and Superconformal Models'}, Nucl. Phys. {\bf B296} (1988) 757.}

\lref\mtwo{E. Bergshoeff, E. Sezgin and P. K. Townsend,
 {\sl `Supermembranes and Eleven Dimensional Supergravity'}, Phys.
Lett. {\bf B189} (1987) 75\semi M. J. Duff and K. S. Stelle, {\sl
`Multimembrane solutions of D=11 Supergravity'}, Phys. Lett. {\bf
B253} (1991) 113.}
\lref\mfive{R. G\"uven, {\sl `Black $p$--Brane 
Solutions of D=11 Supergravity theory'}, Phys. Lett. {\bf B276} (1992)
49.}

\lref\town{P. Townsend, {\sl `The eleven-dimensional 
supermembrane revisited'},
  Phys. Lett. {\bf B350} (1995) 184, hep-th/9501068}
\lref\goed{E. Witten, {\sl `String Theory Dynamics in Various Dimensions'}, 
Nucl. Phys. {\bf B443} (1995) hep-th/9503124.}
\lref\wznw{S. P. Novikov, Ups. Mat. Nauk. {\bf 37} (1982) 3\semi
E. Witten, {\sl `Non--Abelian Bosonization in Two Dimensions'},
Comm. Math. Phys. {\bf 92} (1984) 455.}
\lref\rohm{R. Rohm, {\sl `Anomalous Interactions for the Supersymmetric 
Non--linear Sigma Model in Two Dimensions'}, Phys. Rev. {\bf D32} (1984) 2849.}

\lref\romans{L. Romans, {\sl `Massive $N{=}2A$ 
Supergravity In Ten-Dimensions'}, Phys. Lett. {\bf B169} (1986) 374.}
\lref\others{E. Bergshoeff, M. de Roo, M. B. Green, G. Papadopoulos
 and P. K. Townsend, 
{\sl `Duality of Type II 7-branes and 8-branes'}, Nucl.Phys. {\bf
B470} (1996) 113, hep-th/9601150.}

\lref\duff{Duff, Minasian and Witten, {\sl `Evidence for Heterotic/Heterotic 
Duality'}, Nucl. Phys. {\bf B465} (1996) 413, hep-th/9601036.}
\lref\duffetal{M. J. Duff, {\sl 
`Strong/Weak Coupling Duality from the Dual String', Nucl. Phys. {\bf B442}
(1995) 47, hep-th/9501030.}\semi M. J. Duff, {\sl `Putting
string/string duality to the test', Nucl. Phys. {\bf B436} (1995) 507,
hep-th/9406198.}}
\lref\berkoozi{M. Berkooz, R. G. Leigh, J. Polchinski, J. Schwarz, N. Seiberg 
and E. Witten, {\sl `Anomalies, Dualities, and Topology of D=6 N=1 Superstring
Vacua'}, Nucl. Phys. {\bf B475} (1996) 115, hep-th/9605184.}
\lref\rozali{M. Rozali,
 {\sl `Matrix Theory and U--Duality in Seven Dimensions'}, 
Phys. Lett. {\bf B400} (1997) 260, hep-th/9702136.}
\lref\berkoozii{M. Berkooz, N. Seiberg and M. Rozali, {\sl 
`Matrix Description of 
M-theory on $T^4$ and $T^5$'}, Phys. Lett. {\bf B408}
 (1997) 105, hep-th/9704089.}
\lref\berkooziii{M. Berkooz, and M. Rozali, 
{\sl `String Dualities from Matrix Theory'}, hep-th/9705175.}
\lref\morten{M. Krogh, {\sl `A Matrix Model for Heterotic
 $Spin(32)/Z_2$ and Type I String Theory'}, hep-th/9801034.}

\lref\atish{A. Dahbolkar and J. Park, {\sl `Strings on Orientifolds'},
 Nucl. Phys.  {\bf B477} (1996) 701, hep-th/9604178.}
\lref\bfss{T. Banks, W. Fischler, S. Shenker and L. Susskind, {\sl 
 `M--Theory As A 
Matrix Model: A Conjecture'}, Phys. Rev. {\bf D55}
 (1997) 5112, hep-th/9610043.}
\lref\edjoe{J. Polchinksi and E. Witten,  {\sl 
`Evidence for Heterotic - Type I 
String Duality'}, Nucl. Phys. {\bf B460} (1996) 525, hep-th/9510169.}
\lref\seiberg{N. Seiberg, {\sl `Matrix Description of M-theory on $T^5$ and 
$T^5/Z_2$'}, Phys. Lett. {\bf B408} (1997) 98, hep-th/9705221.}
\lref\sagnotti{M. Bianchi and A. Sagnotti, {\sl `Twist
 Symmetry and Open String Wilson Lines'} Nucl. Phys. {\bf B361} (1991)
519.}
\lref\sagnottii{A. Sagnotti, {\sl 
`A Note on the Green - Schwarz Mechanism in Open - String Theories'},
Phys. Lett. {\bf B294} (1992) 196, hep-th/9210127.}
\lref\horavawitten{P. Horava and E. Witten, {\sl `Heterotic and Type I String 
Dynamics from Eleven Dimensions'}, Nucl. Phys. {\bf B460} (1996) 506,
hep-th/9510209.}

\lref\dvv{R. Dijkgraaf, E. Verlinde, H. Verlinde, 
{\sl `Matrix String Theory'}, Nucl. Phys. {\bf B500} (1997) 43,
hep-th/9703030.}
\lref\dvvii{R. Dijkgraaf, E. Verlinde, H. Verlinde, {\sl `5D Black Holes and 
Matrix Strings'}, Nucl. Phys. {\bf B506} 121 (1997),  hep-th/9704018.}
\lref\robme{C. V. Johnson and R. C. Myers, 
{\sl `Aspects of Type IIB Theory on Asymptotically Locally Euclidean
Spaces'}, Phys. Rev. {\bf D55} (1997) 6382, hep-th/9610140.}
\lref\douglasi{M. R. Douglas, {\sl `Branes within Branes'}, hep-th/9512077.}

\lref\mumford{D. Mumford and J. Fogarty, {\sl `Geometric Invariant Theory'},
Springer, 1982.}
\lref\edcomm{E. Witten, {\sl `Some Comments On String Dynamics'}, in the 
Proceedings of {\sl Strings 95}, USC, 1995, hep-th/9507121.}

\lref\joetensor{J. Polchinski, {\sl `Tensors From $K3$ Orientifolds'}, 
hep-th/9606165.}
\lref\kronheimer{P. B. Kronheimer, {\sl `The Construction of ALE Spaces as 
Hyper--K\"ahler Quotients'}, J.~Diff. Geom. {\bf 29} (1989) 665.}
\lref\hitchin{N. J.  Hitchin,  {\sl `Polygons and Gravitons'}, Math. Proc. 
Camb. Phil. Soc. {\bf 85} (1979) 465.}
\lref\douglasmoore{M. R. Douglas and G. Moore,
  {\sl `D--Branes, Quivers and ALE Instantons'}, hep-th/9603167.}

\lref\hitchinetal{N. J. Hitchin, A. Karlhede, U. Lindstr\"om and M. Ro\u cek, 
{\sl `Hyper--K\"ahler Metrics and Supersymmetry'}, Comm. Math. Phys. {\bf 108}
(1987) 535.}

\lref\klein{F. Klein, {\sl `Vorlesungen \"Uber das Ikosaeder und die 
Aufl\"osung der Gleichungen vom f\"unften Grade'}, Teubner, Leipzig 1884;
F. Klein, {\sl `Lectures on the Icosahedron and  the Solution of an Equation 
of Fifth Degree'}, Dover, New York, 1913.}

\lref\elliot{J. P. Elliot and P. G. Dawber, {\sl `Symmetry in Physics'}, 
McMillan, 1986.}

\lref\ericjoe{E. G. Gimon and J. Polchinski, {\sl `Consistency
 Conditions of Orientifolds and D--Manifolds'}, Phys. Rev. {\bf D54} (1996) 
1667, hep-th/9601038.}
\lref\ericmeI{E. G. Gimon and C. V. Johnson, {\sl `$K3$ Orientifolds'}, Nucl. 
Phys. {\bf B478} (1996), hep-th/9604129.}
\lref\ericmeII{E. G. Gimion and C. V. Johnson, {\sl `Multiple Realisations of 
${\cal N}{=}1$ Vacua in Six Dimensions'}, Nucl. Phys. {\bf B479} (1996), 285,
hep-th/9606176}
\lref\mackay{J. McKay, {\sl `Graphs, Singularties
 and Finite Groups'}, Proc. Symp. Pure. Math. {\bf 37} (1980) 183,
Providence, RI; Amer. Math. Soc.}

\lref\orbifold{L. Dixon, J. Harvey, C. Vafa and E. Witten, {\sl `Strings on 
Orbifolds'}, Nucl. Phys. {\bf B261} (1985) 678;
{\it ibid}, Nucl. Phys. {\bf B274} (1986) 285.}
\lref\algebra{P. Slodowy, {\sl `Simple Singularities and Simple Algebraic 
Groups'}, Lecture Notes in Math., Vol.  {\bf 815}, Springer, Berlin, 1980.}
\lref\gibhawk{G. W. Gibbons and S. W. Hawking, {\sl `Gravitational
Multi--Instantons'}, Phys. Lett. {\bf B78} (1978) 430.}
\lref\eguchihanson{T. Eguchi and A. J. Hanson, {\sl `Asymptotically Flat 
Self--Dual Solutions to Euclidean Gravity'}, Phys. Lett. {\bf B74} (1978) 249.}

\lref\wittenadhm{E. Witten, {\sl `Sigma Models and the ADHM Construction of 
Instantons'}, J.~Geom.  Phys. {\bf 15} (1995) 215, hep-th/9410052.}
\lref\edsmall{E. Witten, {\sl `Small Instantons in String Theory'},  Nucl. 
Phys. {\bf B460} (1996) 541, hep-th/9511030.}
\lref\douglasii{M. R.  Douglas, {\sl `Gauge Fields and D--Branes'},  
hep-th/9604198.}

\lref\phases{E. Witten, {\sl `Phases of $N{=}2$ Theories in Two Dimensions'}, 
Nucl. Phys. {\bf B403} (1993) 159,  hep-th/9301042.}
\lref\edbound{E. Witten, {\sl `Bound States of Strings and $p$--Branes'}, 
Nucl. Phys. {\bf B460} (1996) 335, hep-th/9510135.}

\lref\ADHM{M. F. Atiyah, V. Drinfeld, N. J. Hitchin and Y. I. Manin, {\sl
`Construction of Instantons'} Phys. Lett. {\bf A65} (1978) 185.}

\lref\kronheimernakajima{P. B. Kronheimer and H. Nakajima, {\sl `Yang--Mills 
Instantons on ALE Gravitational Instantons'}, Math. Ann. {\bf 288} (1990) 263.}
\lref\italiansi{M. Bianchi, F. Fucito, G. Rossi, and M. Martinelli, 
{\sl `Explicit Construction of Yang--Mills Instantons on ALE Spaces'}, 
Nucl. Phys. {\bf B473} (1996) 367, hep-th/9601162.}

\lref\italiansii{D. Anselmi, M. Bill\'o, P. Fr\'e, L. Giraradello and A. 
Zaffaroni, {\sl `ALE Manifolds and Conformal Field Theories'},  Int. J. 
Mod. Phys. {\bf A9} (1994) 3007,  hep-th/9304135.}

\lref\nonrenorm{L. Alvarez--Gaume and D. Z. Freedman, {\sl `Geometrical 
structure and Ultraviolet Finiteness in the Supersymmetric Sigma Model'}, 
Comm. Math. Phys. {\bf 80} (1981) 443.} 
\lref\myoldpaper{C. V. Johnson, {\sl `Exact Models of Extremal Dyonic 4D 
Black Hole Solutions of Heterotic String Theory'}, Phys. Rev. {\bf D50} (1994)
4032, hep-th/9403192.}

\lref\gojoe{J. Polchinski, {\sl `Dirichlet Branes and Ramond--Ramond Charges
 in String Theory'}, Phys. Rev. Lett. {\bf 75} (1995) hep-th/9510017.}
\lref\dnotes{J. Polchinski, S. Chaudhuri and C. V. Johnson, {\sl `Notes on 
D--Branes'}, hep-th/9602052.}

\lref\joetasi{J. Polchinski, `TASI Lectures on D-Branes', hep-th/9611050.}
\lref\hull{C. M.  Hull, {\sl `String--String Duality in Ten Dimensions'}, 
 Phys. Lett. {\bf B357} (1995) 545,  hep-th/9506194.}

\lref\dine{M. Dine, N. Seiberg and E. Witten, {\sl 
`Fayet--Iliopolos Terms in String Theory'}, Nucl. Phys. {\bf B289}
(1987) 589.}
\lref\taylor{W. Taylor,
{\sl `D--Brane field theory on compact spaces'}, Phys.Lett. B394
 (1997) 283, hep-th/9611042\semi O. J. Ganor, S. Ramgoolam, W. Taylor,
 {\sl `Branes, Fluxes and Duality in M(atrix)-Theory'},
 Nucl. Phys. {\bf B492} (1997) 191, hep-th/9611202.}

\lref\matrixeight{D. Lowe, {\sl `$E_8{\times}E_8$
 Instantons in Matrix Theory '}, hep-th/9709015\semi O. Aharony,
M. Berkooz, S. Kachru, and E. Silverstein, {\sl `Matrix Description of
$(1,0)$ Theories in Six Dimensions'}, hep-th/9709118.}
\lref\eva{S. Kachru and E. Silverstein, {\sl `On Gauge Bosons in
 the Matrix Model Approach to M~Theory'}, Phys. Lett. {\bf B396} (1997)
70, hep-th/9612162.}
\lref\rey{N. Kim  and S-J. Rey, {\sl `M(atrix) Theory on an Orbifold and 
Twisted Membrane'}, Nucl. Phys. {\bf B504} (1997) 189,
hep-th/9701139\semi S-J. Rey, {\sl `Heterotic M(atrix) Strings and
Their Interactions'}, Nucl. Phys. {\bf B502} 170,1997,
hep-th/9704158.}
\lref\banksmotl{T. Banks and L. Motl, 
{\sl `Heterotic Strings from Matrices'}, JHEP 12 (1997) 004, hep-th/9703218.}
\lref\matrixheterotic{L. Motl, {\sl `Quaternions and M(atrix) Theory
in Spaces with Boundaries'}, hep-th/9612198\semi T. Banks, N. Seiberg,
E. Silverstein, {\sl `Zero and One-dimensional Probes with N=8
Supersymmetry'}, Phys. Lett. {\bf B401} (1997) 30, hep-th/9703052\semi
D. Lowe, {\sl `Bound States of Type I' D-particles
and Enhanced Gauge Symmetry'}, Nucl. Phys. {\bf B501} (1997) 134,
hep-th/9702006\semi D. Lowe, {\sl `Heterotic Matrix String Theory'},
Phys. Lett. {\bf B403} (1997) 243, hep-th/9704041.}
\lref\petr{P. Horava, {\sl `Matrix Theory and
Heterotic Strings on Tori'}, Nucl.  Phys. {\bf B505} 84 (1997),
hep-th/9705055.}
\lref\kabat{D. Kabat and
S-J. Rey, {\sl `Wilson Lines and T-Duality in Heterotic M(atrix)
Theory'}, Nucl. Phys. {\bf B508} 535, (1997) hep-th/9707099.}
\lref\matrixheteroticii{S. Govindarajan, {\sl `Heterotic
M(atrix) theory at generic points in Narain moduli space'},
hep-th/9707164.}
\lref\motl{L. Motl, {\sl `Proposals on Non--Perturbative Superstring
 Interactions'}, hep-th/9701025.}
\lref\banks{T. Banks and N. Seiberg, {\sl `Strings from Matrices'},
 Nucl. Phys. {\bf B497} 41 (1997), hep-th/9702187.}
\lref\danielsson{U. H. Danielsson, G. Ferretti, {\sl `The Heterotic
 Life of the D-particle'}, Int. J. Mod. Phys. {\bf A12} 
(1997) 4581, hep-th/9610082.}
\lref\douglasooguri{M. R. Douglas, H. Ooguri, S. H. Shenker, 
{\sl `Issues in M(atrix) Theory Compactification'}, Phys.Lett. {\bf
 B402} (1997) 36, hep-th/9702203\semi M. R. Douglas, H. Ooguri {\sl
 `Why Matrix Theory is Hard'}, hep-th/9710178.}
\lref\fischler{W. Fischler, A. Rajaraman, 
{\sl `M(atrix) String Theory on K3'},
 hep-th/9704123.}
\lref\edmtheory{E. Witten, {\sl `Five-branes And 
$M$-Theory On An Orbifold'}, Nucl. Phys. {\bf B463} (1996) 383,
hep-th/9512219.}
\lref\sigmamodels{S. J. Gates, C. M. Hull and M. Ro\u{c}ek, 
{\sl `Twisted Multiplets and New Supersymmetric Non--Linear Sigma Models'}, 
Nucl. Phys. 
{\bf B248} (1984) 157\semi M. Ro\u{c}ek, K. Schoutens and A. Sevrin, 
{\sl `Off--Shell WZW Models in Extended Superspace'},
Phys. Lett. {\bf B265} (1991) 303\semi M. Ro\u{c}ek, C. Ahn,
K. Schoutens and A. Sevrin, {\sl `Superspace WZW models and Black
Holes'}, hep-th/9110035, Workshop on String and Related Topics,
Trieste, Itaty, Aug. 8--9 1991. Published in Trieste HEP and Cosmology,
(1991) 995.}
\lref\cft{See for example the wonderful book
 {\sl `Conformal Field Theory'}, P. di Francesco, P.~Matthieu and
 D. S\'en\'echal, Springer, 1997.}
\lref\anatomy{C. V. Johnson, {\sl`Anatomy of a Duality'}, 
Nucl. Phys. {\bf B521} (1998) 71, hep-th/9711082.}
\lref\nickmeal{{\sl `Orientifolds, Branes, and Duality of 4D Gauge Theories'},
 Nick Evans, Clifford V. Johnson and Alfred D. Shapere,
 Nucl. Phys. {\bf B505} (1997) 251, hep-th/9703210. }
\lref\ganor{O. J. Ganor and A. Hanany, 
{\sl `Small $E_8$ Instantons and Tensionless Non--critical Strings'},
Nucl. Phys. {\bf B474} (1996) 122, hep-th/9602120.}
\newsec{Introduction and Summary}
\subsec{Matrix Motivations}
One of the recent satisfying products of the duality industry of the
last three years has been a significant rephrasing of the properties of
string theory. At the very least, we have better understanding of how
to characterize many non--perturbative statements about string theory,
usually using duality to another  theory. Although we have no
proof of duality (in the traditional sense), recasting it in terms of
being a symmetry of a (yet to be fully specified) parent theory,
called ``M--theory'', has been shown to be an extremely economical and
powerful way to proceed.

A partial specification of M--theory has been given\bfss\ in terms of
``Matrix theory'', which captures the physics of certain degrees of
freedom of the theory infinitely boosted in one spatial
direction. While this infinite momentum frame (IMF) definition of the
theory is certainly not the whole story, it has certainly been shown
to be robust, surviving many important tests.

One of these tests is simply to understand whether the various duality
symmetries of string theory can be recovered upon taking suitable
limits. Most of these limits involve compactifying the
matrix theory on manifolds and taking geometrical limits of these
manifolds.  For compactifications to above five dimensions on
manifolds breaking no more than one half of the 32 supersymmetries,
there has been considerable success\refs{\banksreview}\ in reproducing
the known string theories (and some partial success in the case where
a quarter is broken\refs{\anatomy,\rutgers}) and their duality
properties.

To be fair, given that the definition of matrix theory involves some
of the vital ingredients with which we phrase duality at the outset,
we should not be completely surprised to find such a success. There
are precise arguments\refs{\sen,\seibergtwo}\ which explain the
successes of these compactifications (and point to their failure below
six dimensions also), using precisely that fact. Nevertheless,
progress has occurred, because we have been able to restate the duality
results in terms which may generalize beyond the situations in which
we originally discovered them. The understanding of the relevance of
holography; a simple supersymmetric quantum mechanical statement of
the onset of non--commutative geometry at short distance; and the
rephrasing of compactified matrix theory and the resulting string
dualities in terms of properties of field theories, are all elements of
this progress.

Let us turn to the issue of string theory in ten dimensions. We
obtain them from some of the simplest compactifications of M--theory,
and correspondingly, we should get IMF definitions of string theories
by analogous simple compactifications of the matrix theory. These
definitions are called ``matrix string theories'' for obvious reasons.

Although we do not expect any surprises here, there is much to be
gained in this exercise. While obtaining an alternative definition of
weakly coupled string theories from this procedure, (which may or may
not be more useful than the original weakly coupled definitions), we
also have a natural extension of that definition to the theory at not
only very strong coupling, but intermediate coupling as well. (By
contrast, earlier attempts at extending the more standard perturbative
string definition beyond weak coupling were not nearly as successful.)
In fact, we will see that sometimes the description at intermediate
coupling is in some sense the most natural region of coupling space to
which the matrix definition of the string has access. Given that the
original duality statements about the structure of string theory away
from weak coupling concerned very strong coupling, this is also a
bonus of the matrix definition.

Much of the language of the matrix string theory\motl\ technology has
been developed in the context of the type~IIA
string\refs{\banks,\dvv}\ (with extensions to the $E_8{\times}E_8$
heterotic
string\refs{\danielsson,\eva,\banksmotl\matrixheterotic,\rey}), and
less explicitly for the type~IIB string. At weak coupling they are
defined in terms of trivial orbifold fixed point theories in 1+1
dimensions, and the free string Fock space has a description in terms
of winding strings comprising the twisted sectors of the orbifold
theory. The type~IIB string away from weak coupling has also been
described in terms of a fixed point\refs{\sethisuss,\banks}, this time
a 2+1 dimensional interacting one. This is natural, as it really
describes the string at intermediate coupling where the type~IIB
strings are interacting.

We shall begin by reviewing and refining how the matrix strings arise
from the original matrix theory definition. We will then extend the
discussion to the remaining string theories ({\it i.e.,} the $SO(32)$
system), providing a complete description of all five matrix string
theories in ten dimensions. 

\subsec{Summary of Results}

$\bullet$ We observe that the matrix strings at {\sl weak coupling}
are all defined in terms of 1+1 dimensional fixed points similar to
the original type~IIA matrix string theory. (This was already observed
for the $E_8{\times}E_8$ heterotic string.) The basic structure is
simply that the Fock space of the free string is made up of winding
strings of a species which is T--dual to the string in question. While
this was known for the type~IIA and $E_8{\times}E_8$ matrix strings,
we see that it extends to all of the string theories, by simply
following the limits implied by duality and matrix theory. The 1+1
dimensional theories are orbifold conformal field theories.

$\bullet$ The orbifold conformal field theories may each\foot{Here,
our results differ from those presented in ref.\morten\ for the
$SO(32)$ system. We thank T. Banks for pointing out that paper to us
after reading an earlier version of this manuscript.}\ be
characterized as the large $N$ limit of a 1+1
dimensional effective theory defined by a Lagrangian which is of the
(matrix) Green--Schwarz form for the matrix string in question. The
constituent fields are $N{\times}N$ matrices.

$\bullet$ The limits which define the matrix string theories also
define certain supergravity backgrounds, which can be interpreted as
``dual'' descriptions in the sense of ref.\refs{\juan}.  In the free
string limits the supergravity dual is simply the near horizon
geometry of a fundamental string in the supergravity associated to the
T--dual species of string. The infra--red limit of the 1+1 dimensional
field theory defining the free matrix string is associated with the
center of the fundamental string solution. This ``dual'' is therefore
not a good description at the core of the string configuration, as it
breaks down due to strong curvature corrections precisely at this
point, {\it i.e.,} at the infra--red limit\foot{This connection was
made for the type~IIA system in ref.\refs{\juantwo}. Here, we point
out that this behaviour is natural and necessary.}. This behaviour is
expected from string duality, as will be discussed.

$\bullet$ Although the singularity in the supergravity prevents us
from using it as a complete dual definition of the 1+1 dimensional
fixed point, this is not a problem, as the orbifold description is
simple enough to characterize without further appeal to such a
dual. Nevertheless, the supergravity description serves to organize
and inform us about the structure of the matrix definition of the
weakly coupled string, helping to lead to the description of the free
string limit given above.

$\bullet$ Moving away from the weakly coupled limits of the matrix
strings, we find that the supergravity description is
smooth. Especially in the cases which involve ten dimensional
string/string duality (the type~IIB and the $SO(32)$ type~IB/heterotic
pair), there is a complete and concise supergravity dual description
of the space in terms of $AdS_4{\times}S^7$ for the first case, and
$(AdS_4/\IZ_2){\times}S^7$ for the second case. The latter defines a
novel 2+1 dimensional fixed point theory with broken Lorentz
invariance. (Such fixed points were conjectured to exist in
ref.\refs{\petr}. This $AdS_4/\IZ_2$ description is a concrete
proposal for their study.)

In the cases of the type~IIA and $E_8{\times}E_8$ heterotic string
cases, the intermediate and strong coupling situations are best
described in terms of the original $0{+}1$ dimensional matrix system.

$\bullet$ We notice also that the organizing two extra hidden
dimensions of ten dimensional type~IIB string theory, which play a
role in F--theory, appear here in describing the type~IIB matrix
string at intermediate coupling.  That they are non--dynamical (but of
course still important) is seen here to be a consequence of the
holographic nature of the AdS/CFT correspondence of ref.\juan. So of
the apparent twelve dimensions with signature (10,2) involved in
defining non--perturbative type~IIB, a pair of dimensions with
signature (1,1) have no dynamics associated to their size.

The outcome of this investigation is thus a comprehensive
characterization of all of the ten dimensional string theories at all
values of their coupling, in terms of 1+1 and 2+1 dimensional field
theories and quantum mechanics. We find that a supergravity solution
is sometimes a complete dual description, and in all cases they
highlight some of the key features of the matrix theory.  This
framework is appealing\foot{A description of the relationships among
all the strings and their dual theories in terms of field theory fixed
points was anticipated quite a while ago in section~5 of ref.\robme.},
and even though it mainly reproduces much that we already know
(namely, ten dimensional strings and their duals) it may serve as a
vital starting point for defining string theories where we do not have
the usual tools available in ten dimensions. With this in mind, we
close the paper in section~6.2 with some detailed preliminary remarks
concerning such applications.

\newsec{The case of Type~IIB}
The matrix theory definition of M--theory in the infinite momentum
frame (IMF) is given by\bfss\ the ${\cal N}{=}16$ supersymmetric
$U(N)$ quantum mechanics arising from $N$ coincident D0--branes'
world--volume, in the limit $\ell_s{\to}0$ and $N{\to}\infty$. The
special longitudinal direction, $x^{10}$, (initially compactified on a
circle of radius $R_{10}$), is decompactified in the limit also.
The type~IIA string theory used to define this theory has parameters:
\eqn\params{g^{\phantom{.}}_{\rm IIA}=
R_{10}^{3/2}\ell_p^{-{3/2}},\quad\ell_s=\ell_p^{3/2}R_{10}^{-{1/2},} }
where $\ell_s$ is the string length and $\ell_p$ is the eleven
dimensional Planck length.

Consider the matrix definition of (IMF) M--theory compactified on a
torus in the directions $x^8, x^9$. When the torus is small , we
should have a description of the type~IIB string
theory\refs{\paul,\john}\ in the light cone
gauge\refs{\sethisuss,\banks}. T--duality from the D0--brane system
succinctly gives the definition in terms of $N$ D2--branes, on whose
world--volume there lives $2{+}1$ dimensional ${\cal N}{=}8$
supersymmetric $U(N)$ Yang--Mills theory. Representing the torus by a
pair of circles of radius $R_9$ and $R_8$, respectively, the
Yang--Mills coupling is computed as:
\eqn\yangone{{1\over g^2_{\rm YM}} =
 {\ell_s\over {\tilde g}^{\phantom{.}}_{\rm IIA}}={R_8R_9\over
R_{10}}.}  Here, ${\tilde g}^{\phantom{.}}_{\rm IIA}$ is the
$T_{89}$--dual type~IIA string coupling, and the D2--branes are
wrapped on a dual torus (in directions ${\hat x}^8,{\hat x}^9$) of
size ${\tilde R}_8=\ell_s^2/R_8$ and ${\tilde R}_9=\ell_s^2/R_9$.

In the limit where the torus shrinks away ($R_8,R_9{\to}0$), with
$N{\to}\infty$, the dual torus decompactifies, and the strongly
coupled $U(N)$ Yang--Mills theory
flows\refs{\sethisuss,\banks,\seibergthree}\ to a non--trivial
superconformal infra--red fixed point with an $SO(8)$
R--symmetry. This $SO(8)$ is the manifestation\refs{\sethisuss,\banks}
of the spacetime Lorentz symmetry of the lightcone theory thus defined
--- the type~IIB string theory. The coordinates of the ten dimensions
are the manifest $x^1{-}x^7$ and a new dimension ${\hat x}^{10}$, in
which the $SO(8)$ acts, while the direction which goes with time $x^0$
to define the light--cone or IMF directions is a linear combination
of~${\hat x}^8$ and~${\hat x}^9$, set by the ratio
$R_9/R_9{=}g^{\phantom{.}}_{\rm IIB}$, the (matrix) type~IIB string
coupling.

\subsec{The Role of Eleven Dimensional Supergravity}

Notice that equation \yangone\ also tells us that in the limit, the
T--dual type~IIA coupling ${\tilde g}^{\phantom{.}}_{\rm IIA}$ is also
infinite (also, $\ell_s{\to}0$), and we should be working in eleven
dimensional supergravity, with $N$ M2--branes extended in ${\hat x}^8$
and ${\hat x}^9$. In the large $N$ limit, the branes produce a
non--trivial gravitational effect on the spacetime in which they are
embedded, and this is summarized neatly in terms of the M2--brane
supergravity solution. Writing the supergravity solution in the large
$N$ limit in terms of $U{=}r/\ell_s^2$, (where $r$ is the transverse
distance from the core of the brane configuration), defining the
characteristic energy scale of the gauge theory\refs{\juan}, we may
study the renormalization group flow of the theory by moving in the
``near horizon'' spacetime created by the brane configuration.

In the strongly coupled limit ({\it i.e.,} the infra--red, $U{=}0$),
with $N{\to}\infty$, the complete description is in terms of eleven
dimensional supergravity compactified on $AdS_4{\times}S^7$, (with
appropriate choices for the three form potential) where the $S^7$ has
a fixed radius defined in terms of the radius of the $AdS_4$.

This supergravity compatification is conjectured\refs{\juan}\ to be a
complete description of the 2+1 dimensional infra--red fixed point,
because of the following features:

\item\item{{$\bullet$} The curvatures of the compactification are small 
everywhere, and thus supergravity is well--defined.}
\item\item{{$\bullet$} The isometries of  $AdS_4$ form the group $SO(3,2)$, 
which coincides with the superconformal group of the fixed point.}
\item\item{{$\bullet$} The isometries of the $S^7$, the group $SO(8)$, give 
rise to a Kaluza--Klein gauge symmetry in the $AdS_4$ spacetime. This
in turn gives rise to a global $SO(8)$ R--symmetry of the fixed point
theory on the boundary.}

The brane construction suggests that the $2{+}1$ dimensional theory
living on the boundary of $AdS_4$ is the fixed point theory. This
AdS/CFT correspondence is ``holographic'' in the sense that the
physical degrees of freedom of the AdS supergravity can be described
by the theory living on the boundary. This correspondence was made
more precise in refs.\correspond\ where
a precise dictionary between the supergravity/conformal field theory
description was suggested. They gave a precise prescription for the
relation between insertions of operators in the boundary conformal
field theory and supergravity modes in the bulk. In the case in hand,
many entries in the dictionary were verified explicitly in
ref.\refs{\ofer}.

The matrix definition of the non--perturbative (matrix) type~IIB
string theory may therefore be regarded as having a dual supergravity
description.

We now digress slightly and briefly, to make remarks concerning a
connection to another description of the non--perturbative type~IIB
string.

\subsec{Holography and F--Theory}

The strength of the type~IIB coupling is determined by the shape of
the torus. The complete complex type~IIB coupling is given in terms of
the modular paramater of the torus:
\eqn\iibcoupling{\tau=A^{(0)}+ie^{-\Phi}=A^{(0)}+{i\over 
g^{\phantom{.}}_{\rm IIB}},} where $A^{(0)}$ is the Ramond--Ramond
scalar and $\Phi$ is the dilaton. In the case in hand, we have
$A^{(0)}{=}0$, and $g^{\phantom{.}}_{\rm IIB}{=}R_9/R_8$.

The situation just described above assumed that we had treated both
directions $x^8$ and $x^9$ on the same footing, and so we took
$R_8,R_9{\to}0$ holding fixed the ratio $g^{\phantom{.}}_{\rm
IIB}{=}R_9/R_8{=}1$, {\it i.e.,} $\tau{=}i$. So the $AdS_4{\times}S^7$
limit defines type~IIB at the strong/weak coupling self--dual point.

This description of the type~IIB string at intermediate coupling is
defined globally everywhere, up to an overall $SL(2,\IZ)$
transformation. In general, $\tau$, the complex structure data of the
torus may vary from place to place in the ten dimensional spacetime of
the type~IIB string, giving a description where the string coupling
$\lambda_{\rm IIB}$ varies, with variations in $A^{(0)}$ signaling
the presence of D7--branes and O7--planes and their Hodge
duals\foot{Our description here may be thought of as focussing on a
local piece of such a general type~IIB background, at intermediate
coupling.}.  This is the point of departure for the
F--theory\refs{\vafa}\ description, which describes such vacua of the
type~IIB string in terms of compactifications on elliptically fibred
Calabi--Yau manifolds of a (naively) 12 dimensional theory.
Non--perturbative type~IIB string theory therefore seems to involve
twelve dimensions.

The extra two dimensions of F--theory are, from many points of view,
not on the same footing as the other ten of the type~IIB theory,
however, as they have no independent dynamics associated with their
size. The torus of the extra two dimensions is the memory of the
complex structure of the torus which was shrunken away in coming from
M--theory to type~IIB. The signature of the extra two--space is
apparently $(1,1)$, giving a complete 12 dimensional spacetime with
signature $(10,2)$. The extra two dimensions are regarded as serving
an organisational role in the type~IIB theory.

As we are describing type~IIB non--perturbatively here, we might hope
to see a sign of these extra dimensions and indeed we do: The
$SO(3,2)$ isometry of the $AdS_4$ space, which becomes the
superconformal group of the $2{+}1$ dimensional theory defining the
matrix type~IIB theory is the Lorentz group of flat space with
signature $(3,2)$. This is the natural space in which $AdS_4$ is
defined as a hyperbolic submanifold. Taking the 2+1 dimensional theory
as an auxiliary theory, describing the IIB theory in 9+1 dimensions,
leaves a two dimensional space with signature (1,1) left over. When
combined with the type~IIB's space gives a twelve dimensional
spacetime\foot{Another way to count would be  to simply regard the
defining $AdS_4{\times}S^7$ eleven dimensional supergravity
compactification as intrinsically using a (10,2) space with (1,1)
holographed away in the construction of matrix type~IIB.}\ of signature
(10,2).

The space with signature $(1,1)$ is again non--dynamical, and now we
see why: The extra time--like direction is part of the embedding space
defining the $AdS_4$, while the extra spatial one is ``projected out''
by the AdS/CFT holographic relationship.

A matrix definition of the weakly coupled type~IIB string can be
found by taking the $R_8,R_9{\to}0$ limit, but keeping
$g^{\phantom{.}}_{\rm IIB}{=}R_9/R_8{<<}1$. This will define a 1+1
dimensional fixed point theory very similar to that which defines the
type~IIA case. We will describe that theory first, and return to the
weakly coupled type~IIB string at the end of the next section.

Let us now turn to the case of type~IIA.

\newsec{The case of Type~IIA}

The matrix definition of IMF type IIA string theory arises from that
of the IMF M--theory definition in a way similar to
above\refs{\dvv}. Compactifying on a circle of radius $R_9$ results in
$N$ coincident D1--branes in type~IIB string theory, which have a
$1{+}1$ dimensional $U(N)$ with ${\cal N}=(8,8)$ symmetry. The gauge
coupling is computed to be:
\eqn\yangtwo{{1\over g^2_{\rm YM}}={\ell_s^2\over g^{\phantom{.}}_{\rm IIB}}
=\ell_s^2{R_9\over R_{10}}={\tilde \ell}_s^2 {\tilde g}^2_{IIB}.}
(Here, ${\tilde \ell}_s$ is the S--dual type~IIB string length and
$g^{\phantom{.}}_{\rm IIB}$ is its coupling.)  This theory has an
$SO(8)$ R--symmetry, which are the manifest rotations of the spacetime
transverse to the branes.

The theory has a definition in terms of a ``matrix Green--Schwarz''
action for the type~IIA string\refs{\edjoe,\dvv}:
\eqn\greenschwarz{S={1\over 2\pi}
\int d^2\sigma {\rm Tr}\biggl((D_\mu X^i)^2+\theta^T\gamma^\mu D_\mu\theta
+{\tilde g}^2_{\rm IIA}F^2_{\mu\nu} -{1\over{\tilde g}^2_{\rm
IIA}}[X^i,X^j]^2+{1\over {\tilde g}^{\phantom{.}}_{\rm
IIA}}\theta^T\Gamma_i[X^i,\theta]\biggr).}  The dimensionless coupling
${\tilde g}^{\phantom{.}}_{\rm IIA}$ is the matrix type ~IIA string
coupling, equal to $R_9/\ell_s$. The $X^i$ are eight scalar fields,
and $\theta$ contains two fermionic fields $\theta_L^\alpha$ and
$\theta_R^{\dot\alpha}$ which respectively transform in the ${\bf
8}_v$, ${\bf 8}_s$ and ${\bf 8}_c$ (vector, spinor and conjugate
spinor) representations of the $SO(8)$. They are all $N{\times}N$
hermitian matrices. The world--volume coordinates are
$\sigma^0,\sigma^1$, which are identified with the (rescaled)
spacetime directions $x^0/{\hat R}_9, {\hat x}^9/{\hat R}_9$, so that
$0\leq\sigma\leq2\pi$. The direction ${\hat x}^9$ is a circle of
radius ${\hat R}_9{=}\ell_s^2/R_9.$

In the limit $R_9{\to}0, N{\to}\infty$ defining the matrix type~IIA
string, the theory flows to an infra--red fixed point, which defines
the ``matrix type~IIA string'' with coupling ${\tilde g}^{\phantom{.}}_{\rm
IIA}{=}R_9/\ell_s$. The theory is a trivial orbifold conformal field
theory, based on the sigma model with target space $(R^8)^N/S_N$,
where $S_N$ is the group of permutations of $N$ identical objects (the
D1--branes themselves). The correspondence works roughly as follows:

The $g^{\phantom{.}}_{\rm YM}{\to}\infty$ long distance limit,
defining the infra--red theory has been shown to correspond to the
type~IIA string theory, where the finite length type~IIA strings arise
from the twisted sectors of the orbifold. The $X^i$ represent the
matrix coordinates of the D1--branes.  In the strong coupling limit,
lowest energy configurations are obtained when the matrices commute,
and may be simultaneously diagonalized, up to the action of the Weyl
group, which permutes the eigenvalues of the matrices along the
diagonal.

As one goes once around the world sheet's spatial direction
$\sigma^1{=}{\hat x}^9/{\hat R}_9$, one can come back to the same
configuration up to a permutation of the eigenvalues. One can build up
a closed string of length $n$ by a acting with a permutation involving
$n$ different eigenvalues as one goes around $\sigma^1$, requiring $n$
jumps (windings) of length 1 to return to the starting
eigenvalue. This defines a matrix type~IIA string with momentum
$P_9=n/R_9$. Long strings which survive the limit are those with $n/N$
finite as $N{\to}\infty$.

Notice that the strings which wind to build up the Fock space of the
string in the free limit are actually type~IIB strings. One way to see
this is to notice that the matrix coordinates $X^i$ start out
initially as D1--brane positions, and so those are the strings which
wind, as is manifest from the lagrangian \greenschwarz. But the weakly
coupled matrix type~IIA string occurs when the type~IIB string
coupling is infinite, and so our winding strings are really the
S--dual fundamental type~IIB strings. The supergravity description will
make this explicit too.

In order to describe interactions in the
theory, a twist operator has to be turned on in the theory, which
exchanges eigenvalues at a given point, thus
creating the splitting/joining interaction of the strings.

This interaction vertex is identified with the $\IZ_2$ twist operator
of the conformal field theory. It was shown in ref.\refs{\dvv}\ to
correspond to the type~IIA string vertex. It is the leading irrelevant
operator in the theory and therefore in order to describe the
interaction of the strings, one has to move away from the infra--red
limit. It is also worth noting that as its identification as an
irrelevant operator is consistent with the fact that the target
spacetime singularity $\IR^8/\IZ_2$ is not able to be smoothly resolved
by switching it on.

\subsec{The Role of Type~IIB Supergravity}

As the theory involves a large number ($N$) of D1--branes, in the
limit $\ell_s{\to}0$, we may also consider the supergravity fields
created by them, in an analogous fashion to the case of defining
the type~IIB string in the previous section.

Again, $U=r/\ell_s^2$ defines an energy scale in the theory, and the
solution may be rewritten in terms of this coordinate. We may study
the renomalization group flow of the theory in these terms. The $N$
D1--brane solution was written in these coordinates in
ref.\refs{\juantwo}. Its behaviour is (neglecting many constants for
clarity):
\eqn\Donebrane{\eqalign{ds^2&=\ell_s^2\left({U^3\over 
g^{\phantom{.}}_{\rm YM}\sqrt{N}}
(dx_0^2+dx_1^2)+{g^{\phantom{.}}_{\rm YM}\sqrt{N}\over
U^3}dU^2+g^{\phantom{.}}_{\rm YM} {\sqrt{N}\over
U}d\Omega_7^2\right)\cr e^\Phi&=\left(g_{\rm YM}^6 N\over
U^6\right)^{1\over2}.}}

In the low energy limit ($U{\to}0$), we approach the core of the
configuration where we see that the dilaton (and hence the type~IIB
string coupling) grows large, infinite in the limit. This is
consistent with the field theory analysis above. We use S--duality to
transform to a solution where the coupling is small in this region,
giving the fundamental string solution:
\eqn\Fonebrane{\eqalign{ds^2&={\hat \ell}_s^2
\left({U^6\over g^4_{\rm YM}{N}}
(dx_0^2+dx_1^2)+{1\over g^2_{\rm YM}}dU^2+
 {U^2\over g^2_{\rm YM}}d\Omega_7^2\right)\cr e^\Phi&=\left(g_{\rm YM}^6
 N\over U^6\right)^{-{1\over2}}.}}

The string coupling vanishes at the core ($U{=}0$) of this $N$
fundamental IIB string configuration. The curvature diverges there,
however, signaling that the IIB supergravity breaks down, just
as we approach the infra--red limit, as already observed in
ref.\juantwo.

Let us further remark here that this supergravity analysis is
perfectly consistent with the matrix string discussion recalled above:

\item\item{{$\bullet$} The strong coupling limit of the Yang--Mills theory 
is also the strong coupling of the type~IIB theory, turning the
D1--branes into F1--branes (fundamental strings). This occurs here in
the same coupling/energy regime. }
\item\item{{$\bullet$} We learned from the field theory analysis that the 
moduli (target) space contains unresolvable orbifold
singularities. Because of the self--duality of the type~IIB theory, we
should take the 1+1 dimensional D1--brane field theory lessons
seriously for the dual fundamental string also\robme. The curvature
divergence at the core of the fundamental string configuration is the
supergravity realization of this phenomenon.}
\item\item{{$\bullet$}  The free string theory is singular from the 
supergravity perspective. The region where supergravity is valid is
away from the infra--red, where the vertex operator representing the
string coupling is switched on. So supergravity can be used to give a
definition of the matrix type~IIA string only away from weak
coupling.}

\subsec{The Weakly Coupled Type~IIB String}
  
In the case of the type~IIB matrix string definition of the previous
section, eleven dimensional supergravity gave the defining infra--red
theory in terms of a compactification on $AdS_4{\times}S^7$. In the
limit, we took $R_8{\sim}R_9{\to}0$ and therefore we have also defined
type~IIB at intermediate coupling but at a non--trivial infra--red
fixed point, by contrast.

We may define a weakly coupled limit of the matrix type~IIB string by
taking the $R_8,R_9{\to}0$ limit, but keeping $g^{\phantom{.}}_{\rm
IIB}{=}R_9/R_8{<<}1$, (or its inverse). In this case, the ${\hat x}^9$
direction of our M--theory configuration effectively shrinks away,
taking us back to ten dimensional type~IIA supergravity (see next
subsection). The 2+1 dimensional fixed point theory under discussion
becomes effectively 1+1 dimensional\refs{\banks}. It is a 1+1
dimensional fixed point. One might imagine that it is essentially an
orbifold theory.  This theory must clearly (see the next subsection
for confirmation by the supergravity dual) be the matrix
Green--Schwarz action
\greenschwarz\ (at strong coupling)
 where now the Green--Schwarz fermions have the same chirality. This
is the type~IIB Green--Schwarz action.

This is the analogue of the matrix type~IIA theory, where now the
winding strings which make up the Fock space of the weakly coupled
matrix type~IIB string are fundamental type~IIA strings.

\subsec{The Role of Type~IIA Supergravity}

It is easy to see that the supergravity limit bears witness to this
description also: the M2--branes wrap the ${\hat x}^9$ circle as it
shrinks away and become $N$ fundamental type~IIA strings lying along
the ${\hat x}^8$ direction. We therefore have a description of the
theory in terms of the neighbourhood of the core of the $N$
fundamental type~IIA string solution in type~IIA supergravity.  These
are the strings which wind and make up the Fock space of the matrix
type~IIB string. This solution is singular at the core again. The
singularity occurs just as we get to the 1+1 dimensional fixed point.

\newsec{The $E_8{\times}E_8$ Heterotic String.}

Thus encouraged by the above complementary pictures, showing us how to
define the type~II matrix string theories, we should expect a
sharpening of the matrix string definitions of all of the remaining
string theories in ten dimensions.

The $E_8{\times}E_8$ heterotic string arises from placing M--theory on
a line interval in $x^9$, say. A matrix definition of the heterotic
string\refs{\danielsson,\eva,\banksmotl\matrixheterotic,\rey}\
proceeds by using the type~IA string theory background, working with
the quantum mechanics of $N$ D0--branes in the presence of a
collection of 16 D8--branes with two orientifold O8--planes, a
distance $\pi R_9$ apart.  Eight of the branes are at each orientifold
plane. (The matrix $E_8{\times}E_8$ heterotic string coupling is
$g^{\phantom{.}}_{\rm HA}{=}R_9/\ell_s$.)

The $0+1$ dimensional model has an $O(N)$ gauge symmetry with an
$SO(16){\times}SO(16)$ global symmetry coming from the background
branes. Bound states of the D0--brane system localized on each 8
D8--brane + 1 O8--plane ``wall'' correspond to spacetime vectors carrying
that gauge symmetry.  In the limit, each $SO(16)$ is filled out to
$E_8$ by 128 additional bound states (a spinor of $SO(16)$) becoming
massless\eva.

The matrix definition of the lightcone $E_8{\times}E_8$ heterotic
string should arise in the limit $R_9{\to}0$. As before, the
description is given in terms of the $T_9$--dual system, which is in
this case the type~IB $SO(32)$ system. The $N$ D0--branes turn into
$N$ D1--branes lying along the direction ${\hat x}^9$ (with radius
${\hat R}_9$), while the 16 D8--branes + 2 O8--planes turn into the 16
D9--branes, a pure world sheet parity projection $\Omega$, with a
Wilson line breaking the $SO(32)$ to $SO(16){\times}SO(16)$.

The $1{+}1$ dimensional $O(N)$ Yang--Mills theory has $(0,8)$
supersymmetry and a coupling:
\eqn\anothercoupling{{1\over g_{\rm YM}^2}=\ell_s^2{R_9\over R_{10}}
={\ell_s^2\over g^{\phantom{.}}_{\rm IB}}={\hat\ell}_s^2g^2_{\rm HB}.}
(Here ${\hat\ell}_s$ is the heterotic string length and
$g^{\phantom{.}}_{\rm HB}$ is the $SO(32)$ heterotic string coupling.)
Once again, there is an $SO(8)$ R--symmetry coming from the rotations
transverse to the D1--branes. The defining action may be thought of a
``matrix Green--Schwarz'' action for the heterotic string:
\eqn\greenschwarzhet{S={1\over 2\pi}
\int d^2\sigma {\rm Tr}\biggl((D_\mu X^i)^2+\theta^T \gamma^\mu 
D_\mu\theta + g^2_{\rm HA}F^2_{\mu\nu}-{1\over g_{\rm
HA}^2}[X^i,X^j]^2+{1\over g^{\phantom{.}}_{\rm
HA}}\theta^T\Gamma_i[X^i,\theta]+\chi^AD_L\chi^A\biggr).}  
The fermion
$\theta$ now contains two fermions $\theta^\alpha_L,
\theta^{\dot\alpha}_R$ which are in the ${\bf 8_s}, {\bf 8_c}$. The
$\theta_L$ are the superpartners of the world sheet gauge field. The
$\chi^A$ are 32 real fermions coming from strings stretched between
the D9--branes and the D1--branes. The effect of the Wilson line is to
make 16 of them periodic and the other 16 antiperiodic. As before
$R_9{\to}0,N{\to}\infty$ defines a weakly coupled matrix string
theory, this time of the $E_8{\times}E_8$ heterotic string with
coupling $g^{\phantom{.}}_{\rm HA}{=}R_9/\ell_s$, as a trivial
orbifold conformal field theory limit.  The orbifold moduli (target)
space obtained in the strong (field theory) coupling limit
is\refs{\banksmotl}\
\eqn\target{{\cal M}={(\IR^8)^N\over S_N\times (\IZ_2)^N}.}
where  and the $\IZ_2$ acts on the 32
current algebra fermions $\chi^A$.

As the type~IB coupling is large here, the winding strings which
build up the Fock space of the $E_8{\times}E_8$ heterotic string are
$SO(32)$ heterotic strings (with the Wilson line) in complete analogy
with the matrix type~IIA case.

The GSO projection assembles $E_8{\times}E_8$ from the
$SO(16){\times}SO(16)$ states in the usual way, combining the $\bf
(1,120){\oplus}(120,1)$ from the periodic--periodic sector with the
$\bf (1,128){\oplus}(128,1)$ from the periodic--antiperiodic sectors,
while throwing out the $\bf (16,16)$. That these are the sectors which
survive at large $N$ should be enforced by the fact that the
expression for the momentum of the heterotic strings in the $x^9$
direction is shifted away from the naive value by the presence of the
Wilson line\eva.

\subsec{The Role of ${\cal N}{=}1$ Supergravity + $SO(32)$ Yang--Mills}

The supergravity discussion is similar to that for the matrix type~IIA
case, now using D1--brane and F1--brane solutions of the $D{=}10,{\cal
N}{=}1$ supergravity. Of course, this theory is anomalous, and this is
cured by adding the $SO(32)$ $D{=}10, {\cal N}{=}1$ super--Yang--Mills
theory to it\foot{There is of course another choice to fix the
anomaly, the system with gauge symmetry $E_8{\times}E_8$. As one might
expect, we will see this arise when we consider the matrix theory of
the weakly coupled $SO(32)$ heterotic string.}. Type~IB/heterotic
duality will also come into play at the level of supergravity to
exchange the $N$ D1--branes into $N$ winding F1--branes as one
approaches the core of the solution. These are the winding $SO(32)$
heterotic strings. Again, at the core, the singularity signals the
approach of the free 1+1 fixed point.

\newsec{The $SO(32)$ Heterotic/Type~IB case.}

Turning to the $SO(32)$ system, we know that this should be realized
by compactifying M--theory on a cylinder.  In other words, we must
compactify the system of the previous section on an additional circle,
say in the $x^8$ direction, of radius $R_8$. We must then take
$R_8,R_9{\to}0$ to find a ten--dimensional theory.

It is easy to see that we recover at finite (but small) $R_8,R_9$ a
description in terms of a D2--brane stretched between two copies of
the 8 D8 + 1 O8 system, now pointlike in ${\hat x}^8$, a distance
$\pi{\tilde R}_8$ apart, where ${\tilde R}_8=\ell_s^2/R_8$. We are in
the type~IA system again, with coupling
\eqn\couple{{1\over g_{\rm YM}^2}={\ell_s\over
 {\tilde g}^{\phantom{.}}_{\rm IIA}} ={R_8R_9\over R_{10}}.}

In taking $R_8,R_9{\to}0$ we will approach the strong type~IA coupling
limit. This time, with the $SO(16){\times}SO(16)$ Wilson line, there
is no choice\refs{\edjoe,\horavawitten,\dnotes,\anatomy}\ but for the
system to go to M--theory on a line interval, the D2--branes becoming
M2--branes (extended in ${\hat x}^8, {\hat x}^9$) while the 8+1
dimensional system at each end of the boundary each become a single
``M9--plane'' defining the ends of the M--theory line interval in
${\hat x}^8$.

Notice that we recover the full $SO(8)$ rotations of the transverse
directions, in this limit, as we have gained the extra direction
$x^{10}$. This translates into the R--symmetry of the field theory on
the M2--brane, which in turn corresponds to the Lorentz group of the
$SO(32)$ matrix string system in the IMF.

There are three distinct versions of the $R_8,R_9{\to}0$ limit:

\subsec{${R_9/R_8}{<<}1.$  The Weakly Coupled $SO(32)$ Heterotic String.}

Generically, the system defines an effective 1+1 dimensional system,
because the ${\hat x}^9$ direction decompactifies faster than the
interval in ${\hat x}^8$. We have an effective dimensional reduction
of the theory on the M2--branes, as one of their directions is
stretched between the M9--planes. Before taking the
limit, the type~IA system of D2--branes stretched between the ends of
the interval give an $O(N)$ Yang--Mills theory with coupling
\eqn\couples{{1\over g_{\rm YM}^2}=
{{\tilde R}_8\ell_s\over{\tilde g}^{\phantom{.}}_{\rm IIA}}= {{\tilde
R}_8R_9R_8\over R_{10}} ={\ell_s^2R_9\over R_{10}.}}  This is a 1+1
dimensional analogue of the gauge theory constructions of Hanany and
Witten\refs{\amied}, which were generalized to include orthogonal and
symplectic groups by adding orientifolds in ref.\refs{\nickmeal}.

In the full limit therefore, we get a 1+1 dimensional theory which
descends from stretching the M2--branes between M9--planes and taking
the limit as the planes approach one another. This defines a 1+1 fixed
point theory with ${\cal N}{=}(0,8)$ supersymmetry and the required
$SO(8)$ R--symmetry coming from rotations in $x^1{-}x^7,x^{10}$.

This is precisely the reduction we want to describe a weakly coupled
matrix SO(32) heterotic string. The $SO(32)$ heterotic string coupling
is proportional to the size of the original interval, as it should be:
\eqn\stringcoupling{g^{\phantom{.}}_{\rm HB}={R_9\over R_8}=
{{\tilde R}_8\over{\tilde R}_9}<<1.}
 
Once again, we can think of this effective theory as arising from the
flow from an effective 1+1 dimensional Yang--Mills theory with $SO(8)$
R--symmetry. We can say precisely what the content of this 1+1
dimensional theory must be. It is of course the matrix Green--Schwarz
action for the $SO(32)$ heterotic string, which is the same action as
eqn.\greenschwarzhet, the same effective action as found on the
D1--brane in type~IB above, but now the winding $SO(32)$ heterotic
strings which built up the Fock space of that theory are replaced by
winding $E_8{\times}E_8$ heterotic strings, as dictated by the
limits. Now we have the other GSO projection on the 32 fermions, which
throws away the spinors of $SO(16)$ and allows the vectors $\bf (16,16)$
to join the $\bf (1,120){\oplus}(120,1)$, filling out the adjoint of
$SO(32)$.

A quick way to see this is to realize that in this limit we have
pushed the D8--branes together fast enough that we never need to
approach the M--theory limit (see ref.\anatomy\ for a relevant
discussion). T--dualizing along the small ${\hat x}^8$ direction, we
can work in terms of $N$ D1--branes in type~IB, with the full $SO(32)$
restored: we have recovered the correct GSO projection.

This is the exact analogue of that which defined the matrix
$E_8{\times}E_8$ heterotic string, This is the full description of the
matrix $SO(32)$ heterotic string at weak coupling.

\subsec{The Role of ${\cal N}{=}1$ Supergravity + $E_8{\times}E_8$ Yang--Mills}

The supergravity observations that we made earlier can now be used to
lend support to these facts:

The supergravity solution in this limit arises from eleven dimensional
supergravity with $N$ M2--branes stretched between the two M9--planes
at the end of the interval, in the limit where the size of the
interval shrinks and $N{\to}\infty.$ This is best described in terms
of the reduced ten--dimensional theory, which is precisely the
$E_8{\times}E_8$ ${\cal N}{=}1$ super Yang--Mills + ${\cal N}{=}1$
supergravity, as determined by the anomaly considerations of
ref.\refs{\horavawitten}.

So we that the other ${\cal N}{=}1$ supergravity arises naturally in
the story as well, as expected.

The $N$ M2--branes are now effectively $N$ one dimensional objects in
the theory. They are $N$ fundamental $E_8{\times}E_8$ heterotic
strings. We are looking at the core of the fundamental string
solution again. It is singular, signaling the approach of the trivial
fixed point describing the free matrix $SO(32)$ string. These
$E_8{\times}E_8$ fundamental strings build up the Fock space of the
free matrix $SO(32)$ strings.

\subsec{${R_9/R_8}{>>}1.$  The Weakly Coupled $SO(32)$ Type~IB
String. }

To study this limit, we begin again with the M2--branes stretched
between the M9--plane again. This time we see that the ${\hat x}^9$
interval grows more slowly than the ${\hat x}^8$ interval and
therefore we obtain an effective 1+1 dimensional system again. This is
again a 1+1 dimensional fixed point theory, defining the weakly
coupled type~IB $SO(32)$ string, with coupling $g^{\phantom{.}}_{\rm
IB}{=}{R_8/R_9}{=}{{\tilde R}_9/{\tilde R}_8}{<<}1.$

The 1+1 dimensional theory is a $\IZ_2$ orbifold (in ${\hat x}^8$) of
the 1+1 dimensional theory which we found defined the weakly coupled
limit of the type~IIB string. The fixed points of the orbifold are at
infinity. The winding strings which build up the Fock space of the the
matrix type~IB strings are fundamental type~IA strings stretched
between the ends of the ${\hat x}^8$ interval. These are simply
type~IIA strings with endpoints on D8--branes at infinity (in the
limit).

It is easy to see the nature of the effective 1+1 dimensional theory
which flows to the field theory fixed point defining the free
string. It comes from a (matrix) Green--Schwarz action for the
type~IIB string, but {\sl with the extra condition that the strings
ends are fixed at the ends of the ${\hat x}^8$ interval}. This is of
course the definition of the (matrix) type~IB Green--Schwarz string
action!

In contrast to the matrix heterotic string limits, there is no family
of 32 heterotic fermions and so the gauge symmetry must arise
elsewhere. Instead, we have the additional degree of freedom to choose
which D8--brane to end on at each end of the ${\hat x}^8$
interval. This is of course simply the introduction of Chan--Paton
factors! It is clearer to count states working with the covering space
of the ${\hat x}^8$ interval: We have 16 D8--branes plus an
orientifold at each end. We trivially get the adjoint $\bf
(1,120){\oplus}(120,1)$ of the manifest $SO(16){\times}SO(16)$ by
considering each end separately. However, we also must include the
$\bf (16,16)$ coming from considering mixed states. This fills out the
adjoint of $SO(32)$.

The infra--red limit of this $O(N)$ effective 1+1 dimensional matrix
model will define the free matrix $SO(32)$ type~IB string as an
orbifold fixed point. The Fock space is defined by winding type~IA
strings.

\subsec{The Role of Type~IA Supergravity}

The supergravity limits support the conclusions immediately above. In
the limits which we took, the $N$ M2--branes stretched between the
M9--planes become $N$ fundamental type~IA strings stretched between
the two collections of 8 D8--branes + 1 O8--plane at each end of the
${\hat x}^8$ interval.

The supergravity description of the limit is therefore the $N$
fundamental string solution of the type~IIA supergravity, stretched
between two domain walls at infinity. In general, this is the type~IIA
massive supergravity of ref.\romans, but the choice of D8--brane
arrangements we have here sets the cosmological constant to zero. As
there is a one--to--one correspondence between the arrangement of
D8--branes and the value of the cosmological constant, fundamental
type~IIA string solutions in massive type~IIA supergravity with other
choices of the cosmological constant will correspond to the matrix
description of (nearly) free $SO(32)$ type~IB strings with  specific
choices of Wilson line.

(It is amusing to note that in essence, we have constructed a
macroscopic version of the type~I string. It is difficult to construct
it directly as a soliton of the dual $SO(32)$ heterotic string because
it is unstable to breaking into smaller pieces in the ten dimensional
theory. We have evaded that problem here by stretching the string
transverse to the space in which it allowed to break, and sending the
ends off to infinity. In effect, we have magnified the region
``between the D9--branes'' in order to construct a stable $SO(32)$
type~I string.)

\subsec{${R_9/R_8}{=}1.$ The $SO(32)$ System at Intermediate Coupling}

In this case, taking this limit $R_9,R_8{\to}0$ and $N{\to}\infty$
will define an uncompactified M--theory limit, staying in eleven
dimensions. Our matrix string definition is the theory on the
M2--branes, as both spatial directions are on the same footing.

We have therefore a $2+1$ dimensional theory with eight supercharges,
but still possessing $SO(8)$ R--symmetry. It is the large $N$ limit of
a variant of $O(N)$ Yang--Mills theory with coupling
\eqn\coupless{{1\over g_{\rm YM}^2}={\ell_s\over 
{\tilde g}^{\phantom{.}}_{\rm IIA}} ={R_9R_8\over R_{10}}} in the
strong coupling limit.

We expect that this defines a fixed point theory with conformal
invariance. Furthermore, we expect it to define an interacting fixed
point theory, as the matrix strings it defines are not weakly
coupled. This theory has regions where it has 16 supercharges, but
there are two 1+1 dimensional submanifolds with half that
number. There are ``twisted sector'' degrees of freedom living on
those submanifolds. This theory should be an example of an infra--red
limit of the type of orbifold Yang--Mills theories studied in
ref.\refs{\petr,\kabat}. The existence of such a fixed point was
conjectured in ref.\petr.

\subsec{A Return to 11D Supergravity}

The supergravity intuition we have developed over the course of the
paper now helps us again, defining precisely the content of the
theory: The non--trivial superconformal fixed point we require is
defined by  a $\IZ_2$ orbifold of eleven dimensional
supergravity on $AdS_4{\times}S^7$, where the orbifold symmetry acts
on the $AdS_4$ factor, leaving the $S^7$ (and hence the $SO(8)$
R--symmetry necessary for matrix string Lorentz invariance)
untouched. This breaks the rotation symmetry in one of the $AdS$
directions and hence we define a subgroup of the $SO(3,2)$
superconformal symmetry of the 2+1 dimensional fixed point.

The orbifolded $AdS_4$ is described by first placing the ${\hat x}^8$
direction on a circle, identifying ${\hat x}^8$ and $-{\hat x}^8$
(which is clearly a symmetry to begin with) and then decompactifying
the circle again. This results in two 2+1 dimensional boundaries
located at ${\hat x}^8{=}0$ and ${\hat x}^8{=}\infty$. Evidently,
consistency of the theory will require some treatment of these two
boundaries analogous to the treatment of ref.\refs{\horavawitten}\ for
the $E_8{\times}E_8$ heterotic string dual. The $\IZ_2$ action will
also act on the fields in the supergravity, and throwing out those odd
(and presumably adding appropriate twisted sectors at the boundary)
will define via the AdS/CFT correspondence, the spectrum of the 2+1
fixed point theory. The holography will clearly work in both sectors,
projecting the bulk of the orbifold AdS to the bulk of the 2+1
dimensional theory, and the fixed point set will also be projected
onto the fixed point set of the 2+1 theory.

(It is interesting to note that the geometry of the 9+1 dimensional
fixed points, after rescaling using the relation between the ten and
eleven dimensional metrics\refs{\town}, $ds_{11}^2{=}
e^{4\phi/3}\left[(d{\hat x}_4{+}A^\mu
dx_\mu)^2{+}e^{-2\phi}ds_{10}^2\right]$, is precisely the
near--horizon geometry of a fundamental string. ($A$ is the R--R
one--form potential in ten dimensions.))

This orbifold of $AdS$ is an easily stated supergravity prescription,
and therefore merits further study, as a means of describing an
unusual type of fixed point.

\vfill\eject

\newsec{Closing Remarks and Outlook}

\subsec{Big Superstrings}

We have seen that the matrix string description of all of the ten
dimensional string theories is qualitatively the same, {\sl but only
in the neighbourhood of weak coupling}, as one might expect on general
grounds. To summarize:

\item\item{{$\bullet$} The free matrix string  
is defined by a 1+1 dimensional fixed point, which is a trivial
orbifold conformal field theory. The twisted sectors of the orbifold
are made up of winding strings of the T--dual variety. These build the
Fock space of the free matrix string. }

\item\item{{$\bullet$}  In each case,  the
fixed point can be described by  an effective 1+1
dimensional large $N$ gauge theory\foot{Here, our results differ from
those presented in ref.\morten\ for the $SO(32)$ system. We thank
T. Banks for pointing out that paper to us after reading an earlier
version of this manuscript.}.  This theory is a ``matrix
Green--Schwarz'' action, for the string. The fields are $N{\times}N$
matrices. (Precisely at the free string limit, the matrices commute,
and the action decomposes into $N$ copies of the usual Green--Schwarz
action.)}

\item\item{{$\bullet$}  There is a supergravity ``dual'' 
description of the effective gauge theory, which is simply the large
$N$ metric of the fundamental string solution associated to the
T--dual string. (This solution is in the associated supergravity of
the T--dual string, of course). The supergravity description breaks
down at the fixed point, which is associated to the center of the
fundamental string solution which is singular. This singularity is the
dual of the orbifold singularities in the conformal field theory
description.}

Away from the weak coupling limit, the theories are divided into two
classes of behaviour: The type~IIA and $E_8{\times}E_8$ heterotic are
in one class, while the type~IIB and the $SO(32)$ strings are in the
other.

\item\item{{$\bullet$}  For the first class,
 the fixed point describing the matrix string stays 1+1 dimensional
and flows back to the effective gauge theory (it arises from a
description in terms of a D1--brane). By time one gets to infinite
coupling, the most economical description is in terms of matrix
quantum mechanics, as the circle (or line interval) which M--theory
was placed upon decompactifies.}

\item\item{{$\bullet$}  For the second class, 
the fixed point theory grows an extra dimension and becomes 2+1
dimensional (it lives on a M2--brane). The brane unwraps the circle
(or line interval) it was placed upon. As the coupling grows, the
other direction that the brane extends in shrinks, and in the infinite
coupling limit, we get a new effective 1+1 dimensional theory with an
associated trivial fixed point. This is ten dimensional string/string
duality.}

\item\item{{$\bullet$} At intermediate coupling, 
the description is fully 2+1 dimensional. The 2+1 dimensional fixed
point is characterized by an 11 dimensional supergravity description
involving $AdS_4$ (or an orbifold). The latter is a new type of fixed
point which deserves further study.}

This complete picture is very satisfying.

\subsec{Little Superstrings}

Now that we have a complete alternative definition of the familiar
ten--dimensional strings at weak coupling, and their extension to
arbitrary coupling, a next step is obvious.

There have been shown to
exist\refs{\dvvii,\seiberg,\ofereva,\edhiggs}\ consistent string
theories in six dimensions, which previously evaded a direct
construction by the usual ten--dimensional techniques. A matrix string
description may be better suited to characterizing them.  It has been
argued\seiberg\ that all of the (big) superstrings in ten dimensions
have a (little) six dimensional descendant. So far, these strings have
not all been completely described in the matrix manner.

Inspired by the results presented here, we can anticipate some key
feature of the little strings' description:

For the $(0,2)$ six dimensional strings (``type~iia''), a description
is found in terms of a fixed point theory derived from a D1--D5
brane system\refs{\stromingervafa,\dvvii,\douglasetal}. 
This fixed point is  a non--trivial one, in contrast to the ten
dimensional type~IIA case. 

The interpretation of all of this in the present context is that there
should be a smooth supergravity description associated with this fixed
point. Indeed, the fixed point is believed to have a description in
terms of type~IIB supergravity on $AdS_3{\times}S^3{\times}T^4$, as
the brane construction would suggest. This description is smooth. It
describes the iia string system at intermediate coupling, which is
arguably\refs{\seiberg}\ the coupling at which it has its most natural
description.

Notice that like the ten dimensional type~IIB system at intermediate
coupling, this description also has two holographed--out dimensions of
signature $(1,1)$. This might be evidence for the decendant of
F--theory, (``f--theory'') whose existence was suggested by
ref.\refs{\gregsamson}.

For the $(0,1)$ six dimensional strings, one would expect an heterotic
supergravity compactification on $AdS_3{\times}S^3{\times}T^4$ to be
involved in defining the little heterotic systems. Another way to get
$(0,1)$ is of course to compactify further on a manifold which breaks
half of the supersymmetry. This suggests that
$AdS_3{\times}S^3{\times}K3$ might be an alternative
description\foot{This idea arose in a conversation with H. Verlinde.}\
of the little heterotic system. The $E_8{\times}E_8$ intersection
cohomology lattice of $K3$ will give the global symmetry which the
little string is supposed to have. That this is reasonable follows
from (essentially) heterotic/type~IIA duality.

It would be very interesting and useful to determine and classify the matrix
descriptions of all of the little strings along the lines suggested here. 

\bigskip
\bigskip

\bigskip
\bigskip

\noindent
{\bf Acknowledgments:}

\noindent
CVJ was supported in part by family, friends and music. CVJ gratefully
acknowledges numerous useful discussions with Robert C. Myers, and
a useful conversation apiece with Ofer Aharony, Petr Horava and Herman
Verlinde. This research was supported financially by NSF grant
\#PHY97--22022.

\bigskip
\bigskip

\centerline{\epsfxsize1.0in\epsfbox{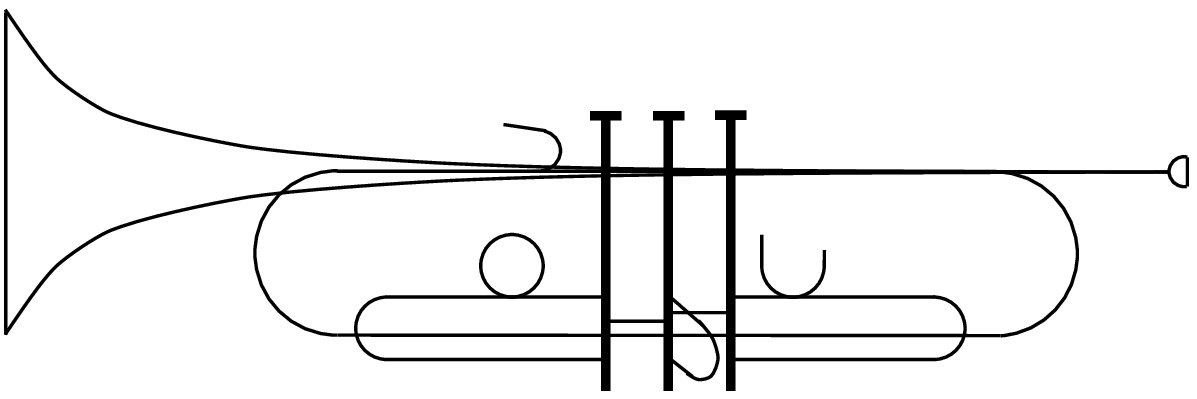}}

\listrefs

\bye